\documentclass[fleqn,twoside,twocolumn,nofootinbib,showkeys]{revtex4} 
\usepackage[sec,nocpr]{ujp} 
\begin{document}
\title[Structure of $^{14}$N Nucleus]
{STRUCTURE OF \boldmath$^{14}$N NUCLEUS\\ WITHIN A
FIVE-CLUSTER MODEL}%

\author{B.E.~Grinyuk, D.V.~Piatnytskyi}
\affiliation{\bitp}
\address{\bitpaddr}
\email{bgrinyuk@bitp.kiev.ua}

\udk{530.143.5} \pacs{27.20.+n, 1.60.Gx,\\[-3pt] 21.10.Ft,
21.10.Gv} \razd{\secii}

\autorcol{B.E.\hspace*{0.7mm}Grinyuk,
D.V.\hspace*{0.7mm}Piatnytskyi}

\setcounter{page}{835}%

\begin{abstract}
The spatial structure of $^{14}$\!N nucleus is studied within a
five-particle model (three $\alpha$-particles plus two
nucleons).\,\,Using the variational approach with Gaussian bases,
the ground-state energy and wave function are calculated for this
five-particle system.\,\,Two spatial configurations in the
ground-state wave function are revealed.\,\,The density
distributions, pair correlation functions, and the momentum
distributions of particles are analyzed and compared with those of
the mirror nuclei $^{14}$\!C and $^{14}$\!O.
\end{abstract}


\maketitle

\section{Introduction}\vspace*{-0.5mm}

In the present paper, we study the structure characteristics of
$^{14}$N nucleus as a system of three $\alpha$-particles and two
extra nucleons (a neutron and a proton).\,\,A steady interest in the
structure of this nucleus can be explained, in particular, by its
important role in the nuclear fusion reactions in stars.

Our five-particle approach may have a rather good accuracy, as it
was shown by calculations of the structure functions of three- and
four-cluster nuclei \cite{R1,R2,R3,R4,R5} consisting of
$\alpha$-particles and two extra nucleons.\,\,The similar
five-particle model \cite{R6} was considered to predict the charge
radius of $^{14}$O nucleus using the clo\-seness of the structures
of mirror nuclei $^{14}$C and $^{14}$O.

The $\alpha$-particle clusters are known to be too tightly bound
systems of four nucleons (with $28.3$ MeV binding energy of $^{4}$He
nucleus) and to have a too small polarizability, so that they can be
considered as structureless particles, as long as one can ignore
their excitation at the impact energy greater than $\sim
$20~MeV.\,\,Although the initial Hamiltonian contains ``pointlike''
$\alpha$-particles, we will consider, after the first-stage
calculations, their size and their own density distributions in the
Helm approximation (see below).\,\,In principle, the nucleon
structure of $\alpha$-particles could be taken into account more
accurately \cite{R1}, if one multiplies the wave function of the
nucleus obtained within the $\alpha$-particle model by the wave
functions of the $^{4}$He nuclei obtained independently in terms of
their nucleon degrees of freedom, and then antisymmetrizes the total
wave function with respect to identical nucleons.\,\,For the ground
state of a nucleus and some low-lying energy levels (for which the
excitation of an $\alpha$-particle can be neglected), our
five-particle model can be competitive in accuracy with the
approaches like \cite{R7}, where one has to deal with all the
nucleon degrees of freedom and thus to resolve a more complicated
problem.

For the five-particle problem, we exploit the variational method
with Gaussian bases \cite{R8,R9} widely used to study the bound
states of few-particle systems.

In the next section, the interaction potentials between particles
are given.\,\,In Section 3, we discuss the r.m.s.\,\,radii and
density distributions of particles in $^{14}$N nucleus.\,\,Relative
distances between particles and pair correlation functions are
presented in Section 4.\,\,Section 5 dwells upon two spatial
configurations in the ground state of $^{14}$N.\,\,In Section 6, the
momentum distributions are given.\,\,Almost in all the cases, we
compare the corresponding structure functions of $^{14}$N nucleus
with those of $^{14}$C and $^{14}$O within the same five-particle
model.

\section{Statement of the Problem}

Within our model, the
five-particle Hamiltonian for $^{14}$N
\[
\hat{H} =
\frac{\mathbf{p}_{1}^{2}}{2m_{p}}+\frac{\mathbf{p}_{2}^{2}}{2m_{n}}+
\sum\limits_{i=3}^{5} \frac{\mathbf{p}_{i}^{2}}{2m_{\alpha}}\,+
\]\vspace*{-7mm}
\[
+\,U_{pn}\left(r_{12}\right)+
\sum\limits_{j>i=3}^{5}\hat{U}_{\alpha\alpha}\left(r_{ij}\right)+
\]\vspace*{-7mm}
\begin{equation}
+\sum\limits_{j=3}^{5}
\hat{U}_{p\alpha}\left(r_{1j}\right)+\sum\limits_{j=3}^{5}
\hat{U}_{n\alpha}\left(r_{2j}\right)+\sum\limits_{j>i=1}^{5}
\frac{Z_{i}Z_{j}e^{2}}{r_{ij}} \label{E1}
\end{equation}
contains, in addition to the kinetic energy, pairwise potentials due
to nuclear and Cou\-lomb interactions between particles.\,\,In
expression (\ref{E1}), the indices $p$, $n$, and $\alpha$ denote a
proton, neutron, and $\alpha$-particle, respectively.\,\,In the
Cou\-lomb term, $Z_{i}$ are the charges of particles in units of
elementary charge $e$: $Z_{1}=1$ for an extra proton, $Z_{2}=0$ for
an extra neutron, and $Z_{3}=Z_{4}=Z_{5}=2$ for $\alpha$-particles.
The nuclear potential $U_{pn}(r_{12})$ between the extra nucleons in
the triplet state is used in the form of a local potential proposed
in \cite{R10} with two Gaus\-sian terms describing the attraction
(with intensity $-146.046$~MeV and radius $1.271$~fm) and the
repulsion (with intensity $840.545$~MeV and radius
$0.44$~fm).\,\,This simple potential gives correct experimental
values for the deuteron binding energy $\varepsilon_{\rm
d}=2.224576$~MeV and charge radius $R_{\rm d}=2.140$~fm, as well as
experimental triplet $np$-scattering length $a_{np,t}=5.424$~fm, and
a good description of the $np$ phase shift in the triplet state (up
to $\sim$$300$~MeV).\,\,This potential was successfully used
\cite{R10,R11,R12} for studying the $^{6}$Li nucleus structure
functions and their \mbox{asymptotics.}

The potentials $U_{n\alpha}$ and $U_{p\alpha}$, as well as the
interaction potential between $\alpha$-particles $U_{\alpha\alpha}$,
are of a generalized type with local and nonlocal (separable)
terms.\,\,This type of potentials was first proposed in
\cite{R13,R14} to simulate the exchange effects between particles in
interacting clusters and was successfully used, in particular, in
calculations \cite{R1,R3,R5,R6} of multicluster
nuclei.\,\,Parameters of the potentials $\hat{U}_{p\alpha}$ and
$\hat{U}_{n\alpha}$, having local attraction and separable
repulsion, are given in \cite{R6}, where these potentials were used
to study the structure of mirror nuclei $^{14}$C and $^{14}$O.\,\,As
for the potential $\hat{U}_{\alpha\alpha}$ between
$\alpha$-particles, its parameters slightly differ from those used
in \cite{R6}.\,\,This little change was necessary to reproduce
accurately the experimental energy and charge radius of $^{14}$N
nucleus.\,\,In the local part of the interaction potential
consisting of two Gaussian terms, we use the same intensity of a
local attraction $-43.5$ MeV and that of a local repulsion 240.0
MeV, but with a little bit enlarged radii: 2.746 fm and 1.530 fm,
respectively.\,\,The separable repulsion of the
$\hat{U}_{\alpha\alpha}$ potential \cite{R6} is not changed.

The ground-state energy and the wave function are calculated with
the use of the variational method in the Gaussian representation
\cite{R8,R9}, which proved its high accuracy in calculations of
few-particle systems.\,\,For the ground state of the five-particle
system (consisting of three $\alpha$-particles plus two additional
nucleons), the wave function can be expressed in the form
\[
\Phi=\hat{S}\sum_{k=1}^{K}C_{k}\varphi_{k}\equiv
\]\vspace*{-7mm}
\begin{equation} \label{E2}
\equiv\hat{S}
\sum_{k=1}^{K}C_{k}\,\exp\left(\!-\sum_{j>i=1}^{5}a_{k,ij}
\left(\mathbf{r}_{i}-\mathbf{r}_{j}\right)^{2}\!\!\right)\!\!,
\end{equation}%
where $\hat{S}$ is the symmetrization operator with respect to the
coordinates of identical $\alpha$-particles, and the linear
coefficients $C_{k}$ and nonlinear parameters $a_{k,ij}$ are
variational parameters.\,\,The greater the dimension $K$ of the
basis, the more accurate the result is obtained.\,\,Note that, at
any $K,$ the trial wave function is exactly invariant with respect
to translations in space, and, thus, the calculated center of mass
kinetic energy is known to be exactly zero.\,\,The linear
coefficients $C_{k}$ can be found within the Galerkin method from
the system of linear equations determining the energy of the system:
\begin{equation} \label{E3}
\sum_{m=1}^{K}C_{m}\left\langle\hat{S}\varphi_{k}
\left|\hat{H}-E\right|\hat{S}\varphi_{m}\right\rangle=0,\,\,\,\,k\,
=\,0,1,...,K.
\end{equation}%
The matrix elements in (\ref{E3}) are known to have explicit form
for potentials like the Coulomb potential or the ones admitting a
Gaussian expansion.\,\,Our potentials between particles just have
the form of a few Gaussian functions, including the Gaussian form
factor in the separable repulsive term.\,\,Thus, system (\ref{E3})
becomes a system of algebraic equations.\,\,We achieved the
necessary high accuracy by using up to $K=600$ functions of the
Gaussian basis.\,\,To fix the nonlinear variational parameters
$a_{k,ij}$, we used both the stochastic approach \cite{R8,R9} and
regular variational methods.\,\,This enables us to obtain the best
accuracy at reasonable values of the dimension $K$.\,\,Note that we
solved, in fact, the five-particle problem a number of times, by
fitting the parameters of the potentials in order to obtain the
experimental binding energy of $^{14}$N nucleus (19.772 MeV
subtracting the own binding energy of $\alpha$-particles) and its
charge radius (2.558 fm \cite{R15}).

As a result of calculations, we have the ground-state wave function
of $^{14}$N nucleus within the five-particle model.\,\,This enables
us to analyze the structure functions of this nucleus.\,\,In the
next section, the density distributions of particles and the charge
density distribution in $^{14}$N are discussed.

\section{Density Distributions\\ and R.M.S.\,\,Radii of \boldmath$^{14}$N Nucleus}

The probability density distribution $n_{i}\left(r\right)$ of the
$i$-th particle in a system of particles with the wave function
$|\Phi\rangle$ is known to be
\begin{equation}
n_{i}\left(r\right)=\left\langle\Phi\right|\delta\left(\mathbf{r}-
\left(\mathbf{r}_{i}-\mathbf{R}_{\mathrm{c.m.}}\right)\right)\left|\Phi\right\rangle\!,
\label{E4}
\end{equation}%
where $\mathbf{R}_{\mathrm{c.m.}}$ gives the location of the center
of mass of the system.\,\,The probability density distributions are
normalized as $\int n_{i}\left(r\right)d\mathbf{r}=1$.

In Fig.~1, we depict the values $r^{2}n_{p}\left(r\right)$,
$r^{2}n_{n}\left(r\right)$, and $r^{2}n_{\alpha}\left(r\right)$,
respectively, for the density distributions (multiplied by $r^{2}$)
of an extra proton, extra neutron, and $\alpha$-particles in
$^{14}$N nucleus.\,\,Note that similar profiles were obtained for
$^{14}$C and $^{14}$O nuclei in \cite{R6}, and this means that
$^{14}$N nucleus may have almost the same structure.\,\,It is
clearly seen that the extra nucleons in such a five-particle nuclei
move mainly inside $^{12}$C cluster formed by
$\alpha$-particles.\,\,The small secondary maximum of curve {\it 1}
at $r\approx 3.4$ fm shows that an extra proton (as well as an extra
neutron) in $^{14}$N nucleus can be found off $^{12}$C cluster, but
with a rather small probability.\,\,We note that an extra proton
appears out of $^{12}$C cluster a little bit more often than an
extra neutron does mainly due to its Coulomb repulsion from the
$\alpha$-particles.\,\,It is demonstrated below that two maxima of
curve \textit{1} (and of the dashed line~\textit{3}) are a
consequence of two spatial configurations distinctly present in
$^{14}$N nucleus.
\begin{figure}%
\vskip1mm
\includegraphics [width=\column] {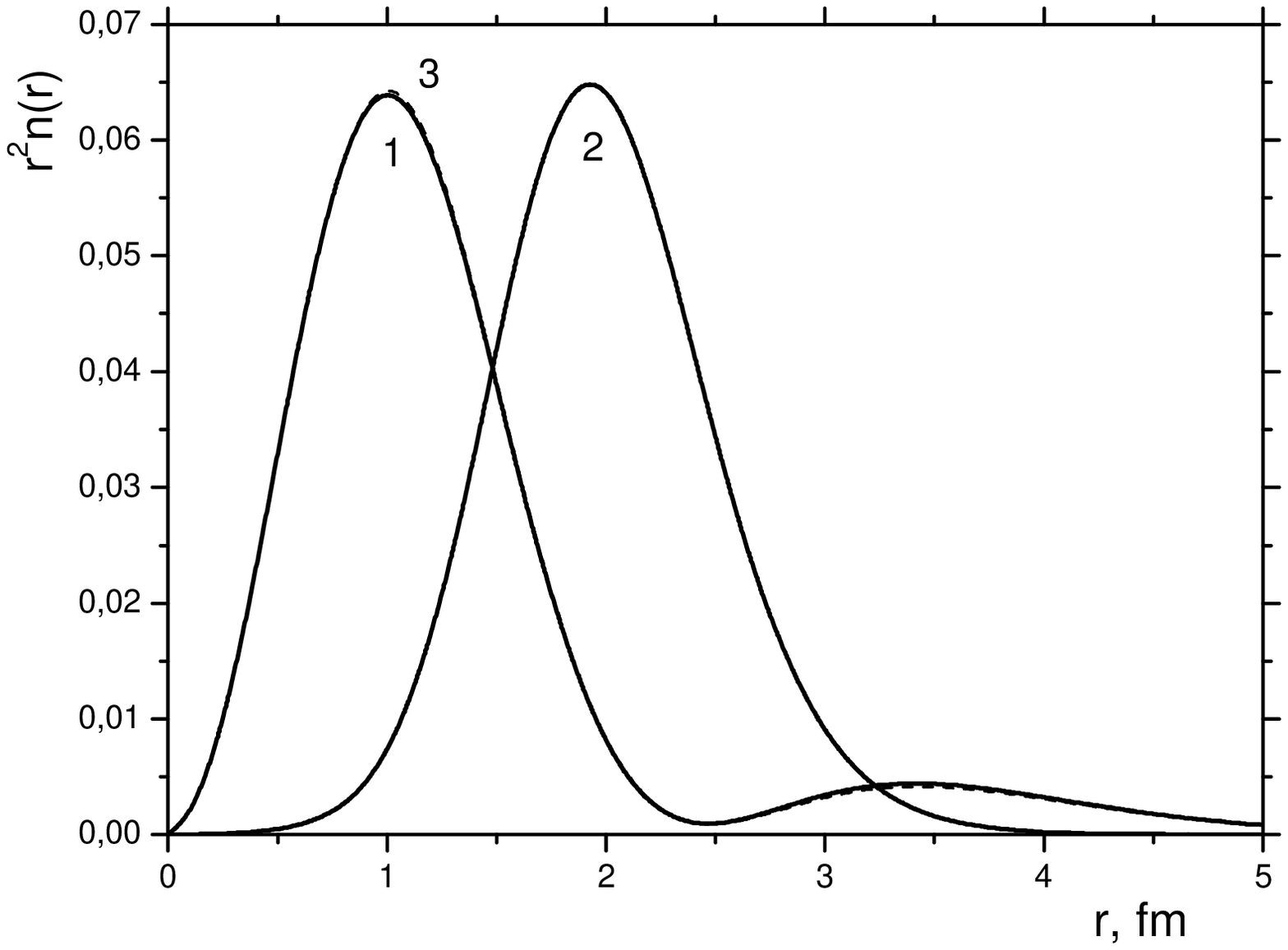}
\vskip-3mm\caption{Probability density distributions multiplied by
$r^{2}$ obtained for an extra proton (solid curve \textit{1}) and
$\alpha$-particles (solid curve \textit{2}) in $^{14}$N nucleus.
Dashed line \textit{3} depicts the same for an extra neutron}
\end{figure}

To find the charge r.m.s. radius of $^{14}$N nucleus, we use the
known Helm approximation \cite{R16,R17}, which enables one, in a
simple way, to take into account that particles are not
``pointlike'' ones.\,\,Within this approach, the charge density
distribution for $^{14}$N nucleus,\vspace*{-3mm}
\[
n_{\mathrm{ch}}\left(r\right)=\frac{6}{7} \int
n_{\alpha}\left(\left|\mathbf{r}-\mathbf{r}'\right|\right)
n_{\mathrm{ch,^{4}He}}\left(r'\right) d\mathbf{r}'+
\]\vspace*{-7mm}
\begin{equation}
+\frac{1}{7} \int
n_{p}\left(\left|\mathbf{r}-\mathbf{r}'\right|\right)
n_{\mathrm{ch,}p}\left(r'\right) d\mathbf{r}'\!, \label{E5}
\end{equation}%
is a sum of convolution products, first being the product of the
density distribution $n_{\alpha}$ for the probability to find an
$\alpha $-particle inside the $^{14}$N nucleus with the charge
density distribution $n_{\mathrm{ch,^{4}He}}$ of an
$\alpha$-particle itself, while the second is the product of similar
distributions for an extra proton.\,\,Coefficients before the
integrals are proportional to the total charge of three
$\alpha$-clusters ($6/7$) and of an extra proton ($1/7$).\,\,The
values of $n_{\alpha}$ and $n_{p}$ are calculated within our
five-particle model according to (\ref{E4}), while
$n_{\mathrm{ch,^{4}He}}$ and $n_{\mathrm{ch,}p}$ follow from the
experimental form factors \cite{R18} and \cite{R19},
respectively.\,\,In relation (\ref{E5}), we neglect the small
contribution of extra neutron.\,\,The normalization of the charge
density distribution is $\int
n_{\mathrm{ch}}\left(r\right)d\mathbf{r}=1$, i.e.\,\,one has to
multiply it by $Ze$ to obtain the necessary dimensional units.

\begin{figure}%
\vskip1mm
\includegraphics [width=\column] {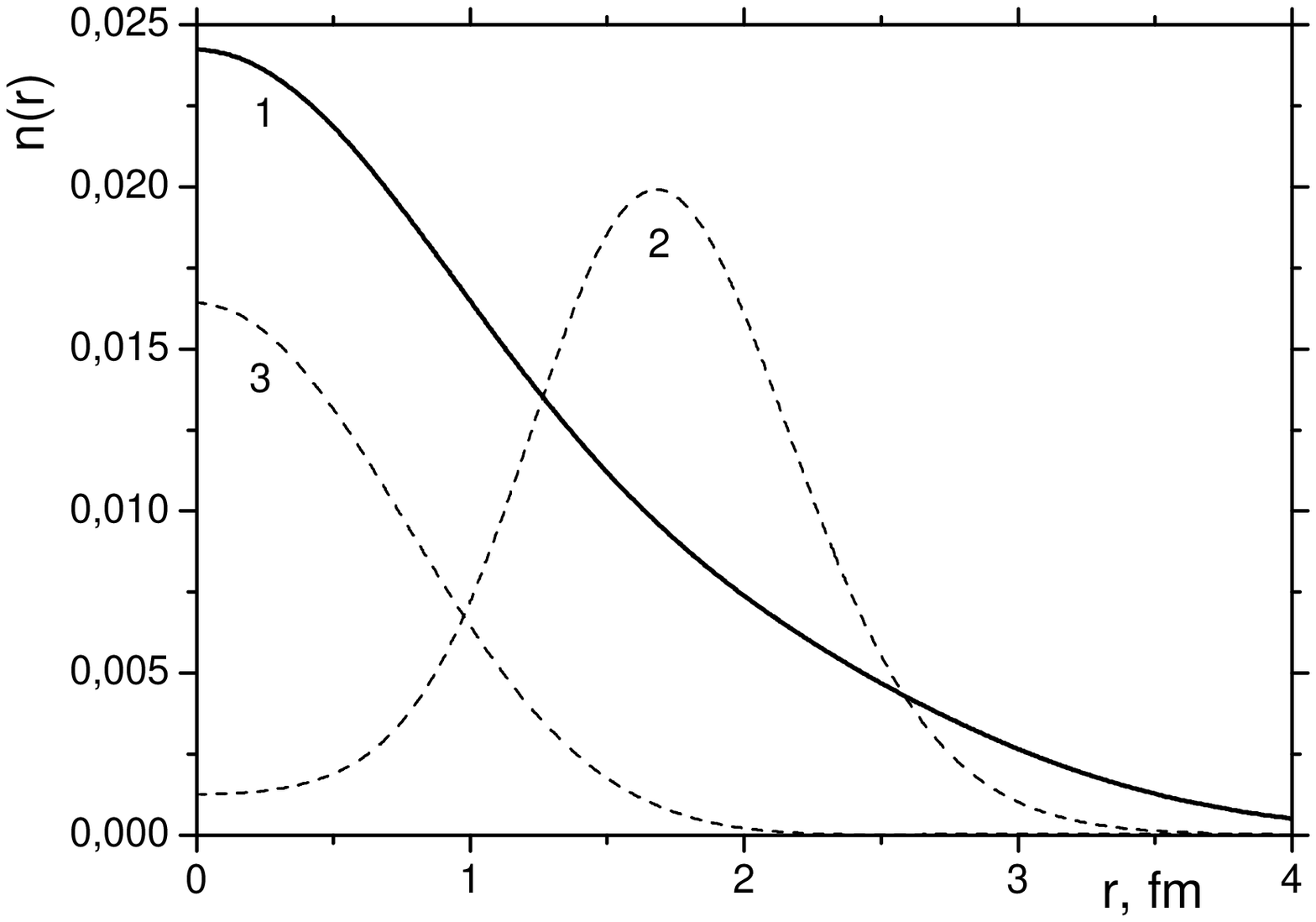}
\vskip-3mm \caption{Charge density distribution in $^{14}$N nucleus
(normalized as $\int n_{\rm ch}\left(r\right)d\mathbf{r}=1$) --
curve \textit{1}. Dashed line \textit{2} depicts the probability
density distribution of ``pointlike'' $\alpha$-particles in $^{14}$N
nucleus. Dashed line \textit{3} is the density distribution
$n_{p}\left(r\right)$ (multiplied by $\times10^{-1}$) of a
``pointlike'' extra proton}
\end{figure}

\begin{table}[b]
\noindent\caption{Calculated r.m.s. relative\\ distances and r.m.s.
radii (fm) for \boldmath$^{14}$N nucleus} \vskip3mm\tabcolsep3.1pt

\noindent{\footnotesize\begin{tabular}{|l|c|c|c|c|c|c|c|c| }
 \hline \multicolumn{1}{|c}
{\rule{0pt}{5mm}$r_{pn}$ }& \multicolumn{1}{|c}{$r_{p\alpha}$} &
\multicolumn{1}{|c}{$r_{n\alpha}$} &
\multicolumn{1}{|c}{$r_{\alpha\alpha}$}&
\multicolumn{1}{|c}{$R_{p}$} & \multicolumn{1}{|c}{$R_{n}$} &
\multicolumn{1}{|c}{$R_ {\alpha}$} & \multicolumn{1}{|c}{$R_ {m}$} &
\multicolumn{1}{|c|}{$R_ {\mathrm{ch}}$}
\\[2mm]
\hline \rule{0pt}{5mm}$2.237$ & $2.692$ & $2.683$ & $3.559$ &$1.598$
& $1.585$ & $2.064$ & $2.556$ & $2.558$ \\[2mm]
\hline
\end{tabular}}
\end{table}

In Fig.~2, the charge density distribution (\ref{E5}) of $^{14}$N
nucleus is shown (solid line \textit{1}).\,\,In spite of the fact
that the density distribution of ``pointlike'' $\alpha$-particles
has a ``dip'' at short distances (see the dashed curve \textit{2}),
its integration with $n_{\mathrm{ch,^{4}He}}$ in (\ref{E5}) smoothes
out this effect completely.\,\,The density distribution of an extra
proton (dashed line \textit{3} depicts $0.1 n_{p}$) makes a little
influence on the total result due to a multiplier $1/7$, but the
proton also contributes at short distances and smoothes the tolal
charge density distribution of the nucleus.\,\,A similar smooth
behavior of the charge density distribution in the Helm
approximation is obtained for $^{14}$C nucleus with two extra
neutrons, and, of course, for $^{14}$O with two extra protons
\cite{R6}.\,\,It is worth to note that the Helm approximation
\cite{R16,R17} used in our model does not involve the exchange
effects between identical nucleons present in the nuclei under
consideration, and this approximation is a rather good one only if
the clusters do not overlap.\,\,To improve the approximation and to
obtain the almost accurate wave function of the nucleus (as noted in
\cite{R1}), one has to multiply the obtained five-cluster wave
function by the wave functions of $\alpha$-particles (expressed in
terms of the nucleon degrees of freedom) and then to carry out the
antisymmetrization of the obtained fourteen-nucleon wave function
over identical nucleons.\,\,This is beyond our study, and thus we
omit a comparison of the results obtained in the Helm approximation
for charge density distributions (and corresponding form factors)
with experimental data.

The r.m.s.\,\,radius $R_{i}$ of a probability density distribution
$n_{i}\left(r\right)$ is known to be $R_{i}=\left(\int
r^{2}n_{i}\left(r\right)d\mathbf{r}\right)^{1/2}$. Having the wave
function in the explicit form of a sum of Gaussian functions, we
obtain the r.m.s.\,\,radii for $^{14}$N nucleus within the
five-particle model.\,\,In Table, the r.m.s.\,\,radii obtained for a
``pointlike'' extra proton $R_{p}$, extra neutron $R_{n}$, and
$\alpha$-particles $R_{\alpha}$ in $^{14}$N nucleus are shown.\,\,We
also give the calculated r.m.s.\,\,matter $R_{m}$ and charge
$R_{\mathrm{ch}}$ radii.\,\,For convenience, we give here also
r.m.s.\,\,relative distances $r_{ij}$ between particles (see the
next section, where the definition of $r_{ij}$ is given, and their
relation to r.m.s.\,\,radii $R_{i}$ is presented).

\section{Pair Correlation\\ Functions and Relative Distances}

More information about the structure of a nucleus can be obtained
from the analysis of the pair correlation functions.\,\,The pair
correlation function $g_{ij}\left(r\right)$ for a pair of particles
$i$ and $j$ can be defined as
\begin{equation}\label{E6}
g_{ij}\left(r\right)=\left\langle\Phi\right|\delta\left(\mathbf{r}-
\left(\mathbf{r}_{i}-\mathbf{r}_{j}\right)\right)
\left|\Phi\right\rangle\!,
\end{equation}
and it is known to be the density of the probability to find the
particles $i$ and $j$ at a definite distance $r$.\,\,These functions
are normalized as $\int g_{ij}\left(r\right)d\mathbf{r}=1$. The
r.m.s.\,\,relative distances squared $\left\langle
r^{2}_{ij}\right\rangle$ are directly expressed through the pair
correlation functions $g_{ij}$:\vspace*{-3mm}
\begin{equation}\label{E7}
\left\langle r^{2}_{ij}\right\rangle=\int r^{2}
g_{ij}\left(r\right)d \mathbf{r}.
\end{equation}
The calculated r.m.s.\,\,relative distances
$r_{ij}\equiv\left\langle r^{2}_{ij}\right\rangle^{1/2}$ between
particles in $^{14}$N nucleus are given in Table.\,\,We note that
the r.m.s.\,\,radii $R_{i}$ are connected with the
r.m.s.\,\,relative distances $r_{jk}$:\vspace*{-2mm}
\begin{equation}\label{E8}
R_{i}^{2}=\frac{1}{M^{2}}\Bigg(\!\!\left(M-m_{i}\right)\sum_{j\neq
i} m_{j}r_{ij}^{2}-\!\!\!\!\!\sum_{\scriptsize\begin{array}{c}
                             j<k\\[-1mm] \left(j\neq i, k\neq
i\right)
\end{array}}m_{j}m_{k}r_{jk}^{2}\!\Bigg)\!,
\end{equation}
where $M$ is the total mass of the system of particles.\,\,Thus, the
r.m.s.\,\,radii $R_{i}$ could be calculated not only directly
through the density distributions, but also (equivalently) with the
use of the pair correlation functions and relations (\ref{E7}) and
(\ref{E8}).\,\,Note that the relative distances between particles
are about (or even a little bit greater) than the sum of their own
sizes.\,\,This fact substantiates, in part, the validity of our
cluster model.

Since the average of a pairwise local potential
$V_{ij}\left(r\right)$ is expressible directly through the pair
correlation function $g_{ij}\left(r\right)$,
\begin{equation}\label{E9}
\left\langle \Phi \right|V_{ij}\left|\Phi\right\rangle=\int
V_{ij}\left(r\right)g_{ij}\left(r\right)d\mathbf{r},
\end{equation}
the variational principle makes the profile of
$g_{ij}\left(r\right)$ such that it has a maximum, where the
potential is attractive, and a minimum in the area of repulsion (if
the role of the kinetic energy is not crucial).\,\,The
$\alpha$-particles have about four times greater mass than extra
nucleons, and, thus, their kinetic energy is essentially smaller
than that of nucleons (see below).\,\,As a result, the pair
correlation function $g_{\alpha \alpha}\left(r\right)$ profile is
determined mainly by the potential $\hat{U}_{\alpha \alpha}$ and has
a pronounced maximum (curve \textit{1} in Fig.~3) near the minimum
of the attraction potential. On the other hand, due to the presence
of a local repulsion in the same potential near the origin, the
profile of $g_{\alpha \alpha}\left(r\right)$ has a dip at short
distances.\,\,Thus, the profile of $g_{\alpha \alpha}\left(r\right)$
shows that $\alpha$-particles are mainly settled at a definite
distance $r_{\alpha\alpha}$ one from another (see Table) being about
the doubled radius of an $\alpha$-particle, and they form a triangle
of $^{12}$C cluster.\,\,The same cluster is present in $^{14}$C and
$^{14}$O nuclei, as seen from Fig.~4, where the pair correlation
functions for $^{14}$C nucleus are shown (we omit almost identical
similar profiles for $^{14}$O).\,\,But since the
$\alpha\alpha$-potential used in the present work has somewhat
greater radius than that accepted in \cite{R6}, it is natural to
obtain $r_{\alpha\alpha}\cong3.6$~fm for $^{14}$N nucleus instead of
$r_{\alpha\alpha}\cong3.2$~fm for $^{14}$C and $^{14}$O nuclei
\cite{R6}.
\begin{figure}%
\vskip1mm
\includegraphics [width=\column] {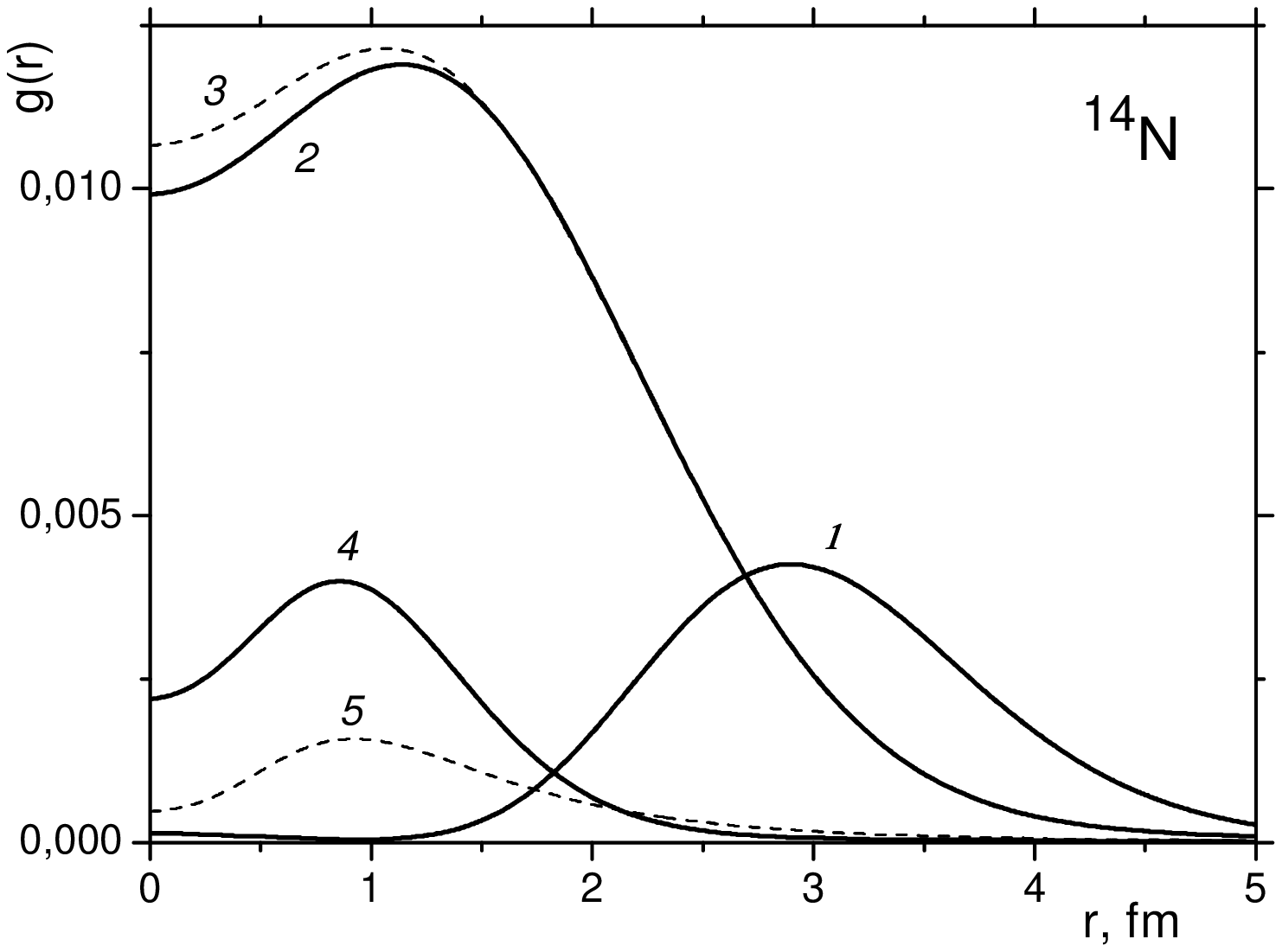}
\vskip-3mm\caption{Pair correlation functions for $^{14}$N nucleus.
Solid line \textit{1} presents $g_{\alpha\alpha}\left(r\right)$,
solid curve \textit{2} depicts $g_{p\alpha}\left(r\right)$, and
dashed line \textit{3} corresponds to $g_{n\alpha}\left(r\right)$.
Curve \textit{4} is the pair correlation function (multiplied by
$0.1$) for extra nucleons, $0.1g_{pn}\left(r\right)$, and dashed
line \textit{5} is the wave function squared of the deuteron
(multiplied by $0.1$)}
\end{figure}

\begin{figure}%
\vskip1mm
\includegraphics [width=\column] {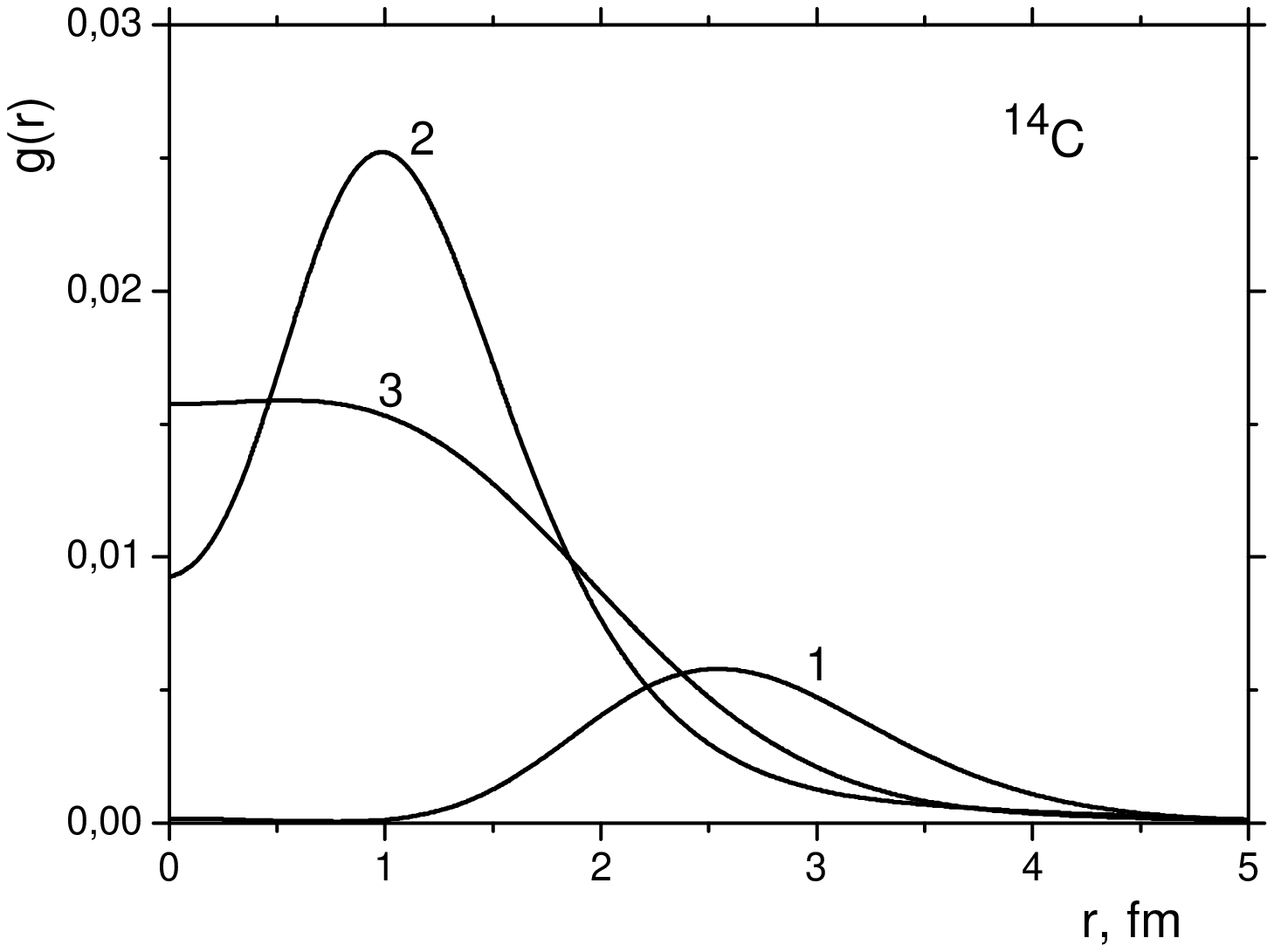}
\vskip-2mm\caption{Pair correlation functions for $^{14}$C nucleus:
$g_{\alpha\alpha}(r)$ -- curve \textit{1}, $g_{nn }(r)$ -- curve
\textit{2}, and $g_{n\alpha}(r)$ -- curve \textit{3}}
\end{figure}

The deuteron cluster in $^{14}$N formed by two extra nucleons has
(on the average, from the qualitative point of view) almost the same
form as a free deuteron, as seen from Fig.~3, where the
$g_{pn}\left(r\right)$ function is shown for $^{14}$N (solid curve
\textit{4}) to be compared with
$g_{pn}\left(r\right)\equiv\left|\psi_{\rm
d}\left(r\right)\right|^{2}$ for a free deuteron (dashed curve
\textit{5}).\,\,But, in $^{14}$N, the deuteron cluster is more
tightly bound than in a free state.\,\,That is why the asymptotics
of the free deutron function $g_{pn}\left(r\right)$ goes above that
of $g_{pn}\left(r\right)$ for $^{14}$N nucleus, while below it at
short distances (due to the normalization condition).\,\,The extra
nucleon pair correlation function ($g_{nn}\left(r\right)$ for
$^{14}$C, as well as $g_{pp}\left(r\right)$ for $^{14}$O), also has
a dip at short distances (see Fig.~4, curve \textit{2}) due to the
presence of a short-range repulsion in our singlet nucleon-nucleon
potential \cite{R3,R5,R6}.

The functions $g_{p\alpha}\left(r\right)$ and
$g_{n\alpha}\left(r\right)$ for $^{14}$N nucleus have a small dip at
short distances (see Fig.~3), while the corresponding functions
$g_{n\alpha}\left(r\right)$ for $^{14}$C and
$g_{p\alpha}\left(r\right)$ for $^{14}$O have no pronounced dips at
all (see Fig.~4 for $^{14}$C).\,\,Almost the same profile is
revealed by $g_{p\alpha}\left(r\right)$ for $^{14}$O, which is not
shown.\,\,The fact that the above-mentioned correlation functions do
not vanish at short distances can be explained by the absence of a
short-range local repulsion in our model of generalized potential
between a nucleon and an $\alpha$-particle.\,\,This potential
contains the local pure attraction plus the nonlocal (separable)
repulsion with greater radius~\cite{R6}.

\section{Two Configurations\\ in \boldmath$^{14}$N, $^{14}$C, and $^{14}$O Nuclei}

To make the structure of the ground state of $^{14}$N nucleus (as
well as of $^{14}$C and $^{14}$O nuclei) more clear, let us
consider the quantity $P\left(r,\rho,\theta\right)$ proportional
to the density of the probability to find extra nucleons at a
definite relative distance $r$ and to find their center of mass at
a distance $\rho$ from the center of mass of $^{12}$C cluster:
\[
P\left(r,\rho,\theta\right)=
\]\vspace*{-9mm}
\begin{equation}\label{E10}
=r^{2}\rho^{2}\left\langle\Phi\right|\delta\left(\mathbf{r}-\mathbf{r}_{NN}\right)
\delta\left(\boldsymbol{\rho}-\boldsymbol{\rho}_{(NN),(3\alpha)}\right)\left|\Phi\right\rangle\!,
\end{equation}
where $\theta$ is the angle between the vectors $\mathbf{r}$ and
$\boldsymbol{\rho}$.\,\,It is assumed that $\theta=0^{\circ}$
corresponds to a spatial configuration, where the extra proton,
extra neutron, and center of mass of $^{12}$C cluster are at the
same line, the proton being further from $^{12}$C than the
neutron.\,\,The angle $\theta=180^{\circ}$ corresponds to almost the
same configuration, but with an extra neutron located further from
the center of mass of $^{12}$C cluster.\,\,If one considers $^{14}$C
and $^{14}$O nuclei, the configurations with $\theta=0^{\circ}$ and
$\theta=180^{\circ}$ are identical due to identical extra nucleons.
Although these two angles are not identical for $^{14}$N nucleus,
the configurations with a definite $\theta$ and $180^{\circ}-\theta$
are very similar (approximately identical), since the role of the
Coulomb interaction is not decisive.\,\,That is why we do not
demonstrate the profiles of $P\left(r,\rho,\theta\right)$ for
$\theta>90^{\circ}$ in the \mbox{figures.}

\begin{figure*}%
\vskip1mm
\includegraphics [width=15.1cm] {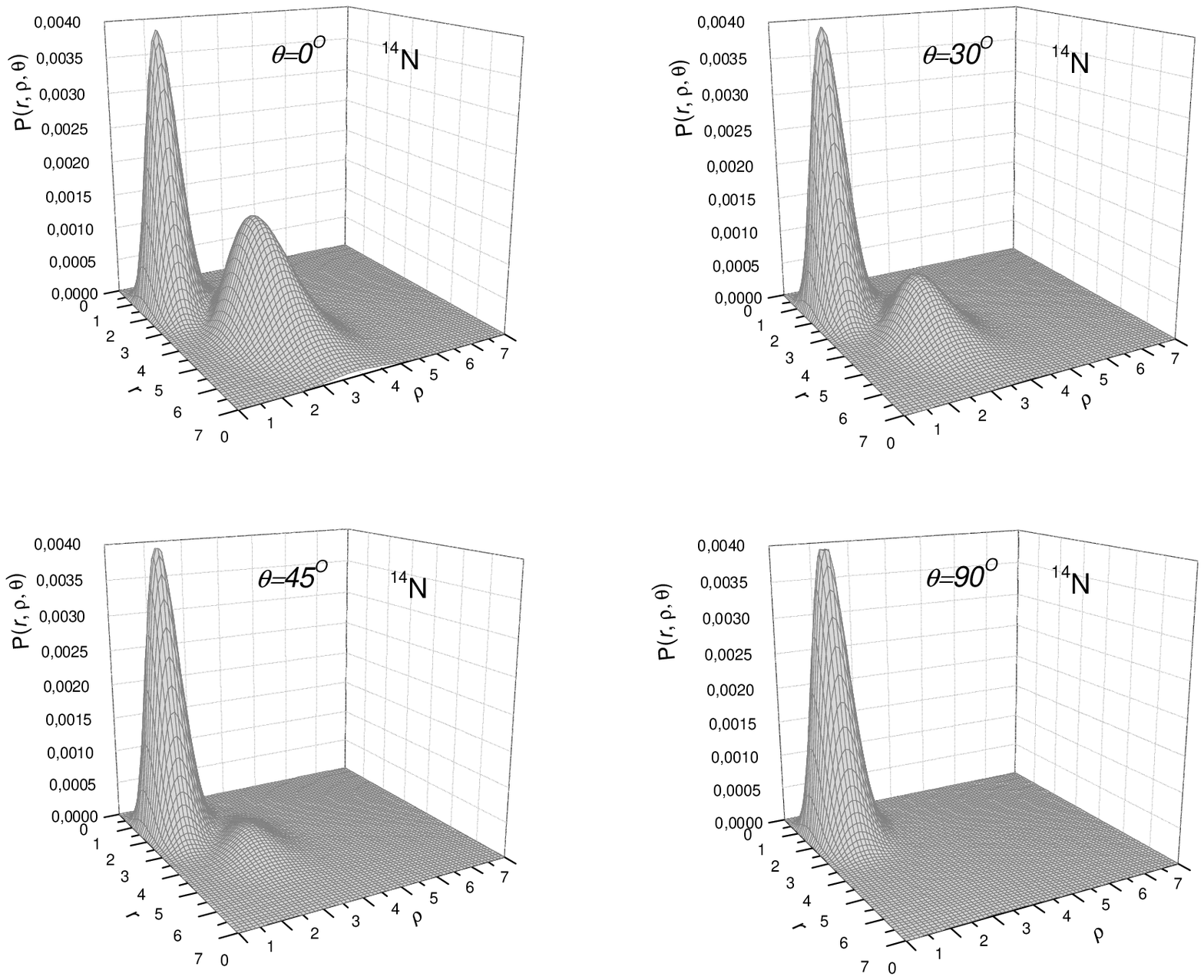}
\vskip-2mm\parbox{15.0cm}{\caption{Two configurations in the ground
state of $^{14}$N nucleus manifesting themselves in the
$P\left(r,\rho,\theta\right)$ function at different
angles~$\theta$}}\vspace*{-1.0mm}
\end{figure*}

The quantity $P\left(r,\rho,\theta\right)$ for $^{14}$N nucleus is
depicted in Fig.~5 for $\theta=0^{\circ}$, $\theta=30^{\circ}$,
$\theta=45^{\circ}$, and $\theta=90^{\circ}$ as a function of $r$
and $\rho$.\,\,Two peaks on the $P\left(r,\rho,\theta\right)$
surface are observed at $\theta=0^{\circ}$ (as well as for
$\theta=180^{\circ}$, which is not shown), and only one peak at
$\theta=90^{\circ}.$ The rest angles give intermediate results (see
Fig.~5 for $\theta=30^{\circ}$ and $\theta=45^{\circ}$).\,\,If it
were not the multiplier $r^{2}\rho^{2}$ in (\ref{E10}), the main
peak present at all the angles $\theta$ would be settled just at
$\rho=0$, i.e.\,\,the center of mass of $^{12}$C cluster and that of
the deuteron one would coincide.\,\,The comparatively smaller (than
a free deuteron, see Fig.~3) deuteron cluster moves mainly inside
$^{12}$C cluster.\,\,The second peak reveals itself mainly at
$\theta=0^{\circ}$ and corresponds to a configuration, where an
extra neutron is located inside $^{12}$C cluster, while an extra
proton is comparatively far from the center of the nucleus (it is
out of $^{12}$C cluster).\,\,At $\theta=180^{\circ}$, almost the
same configuration is observed (not shown in the figure).\,\,But, in
this case, an extra proton is inside $^{12}$C cluster.\,\,Just these
configurations make a contribution to the second maximum of the
extra nucleon probability density distribution (see Fig.~1).\,\,In
this configuration, the center of mass of the subsystem of extra
nucleons does not coincide with the center of mass of $^{12}$C
cluster.\,\,The almost same (from the qualitative point of view) two
configurations are observed in the ground state of $^{14}$C nucleus
(see Fig.~6) or $^{14}$O one (not shown for brevity, since the
corresponding pictures almost coincide with those depicted in
Fig.~6).\,\,Note that a configuration with one nucleon out of
$^{12}$C cluster is more pronounced in the case of mirror nuclei
$^{14}$C and $^{14}$O as compared to $^{14}$N nucleus, because the
interaction potential in the singlet state between extra nucleons is
less strong than the interaction potential in the triplet state,
which compels a proton and a neutron to be coupled inside a
five-particle system with greater probability.\,\,We also note that
a small difference in $\alpha\alpha$-interactions used in
calculations of the $^{14}$N and $^{14}$C nuclei ground states makes
almost no influence on the effect of two configurations.\,\,We
carried out a number of test calculations with
$\alpha\alpha$-potentials, which result in different values of the
binding energy of $^{12}$C nucleus.\,\,The results for the ground
states of $^{14}$N and $^{14}$C nuclei are similar to those shown in
Figs.~5 and~6.\looseness=1

A similar situation with two configurations in the ground state is
found for $^{6}$He, $^{6}$Li \cite{R1,R2,R3,R12} or $^{10}$Be,
$^{10}$C \cite{R4,R5} nuclei, where the center of mass of the
dinucleon subsystem coincides (one configuration) or does not
coincide (another configuration) with the center of mass of the
subsystem of $\alpha$-particles.
\begin{figure*}%
\vskip1mm
\includegraphics [width=15.8cm] {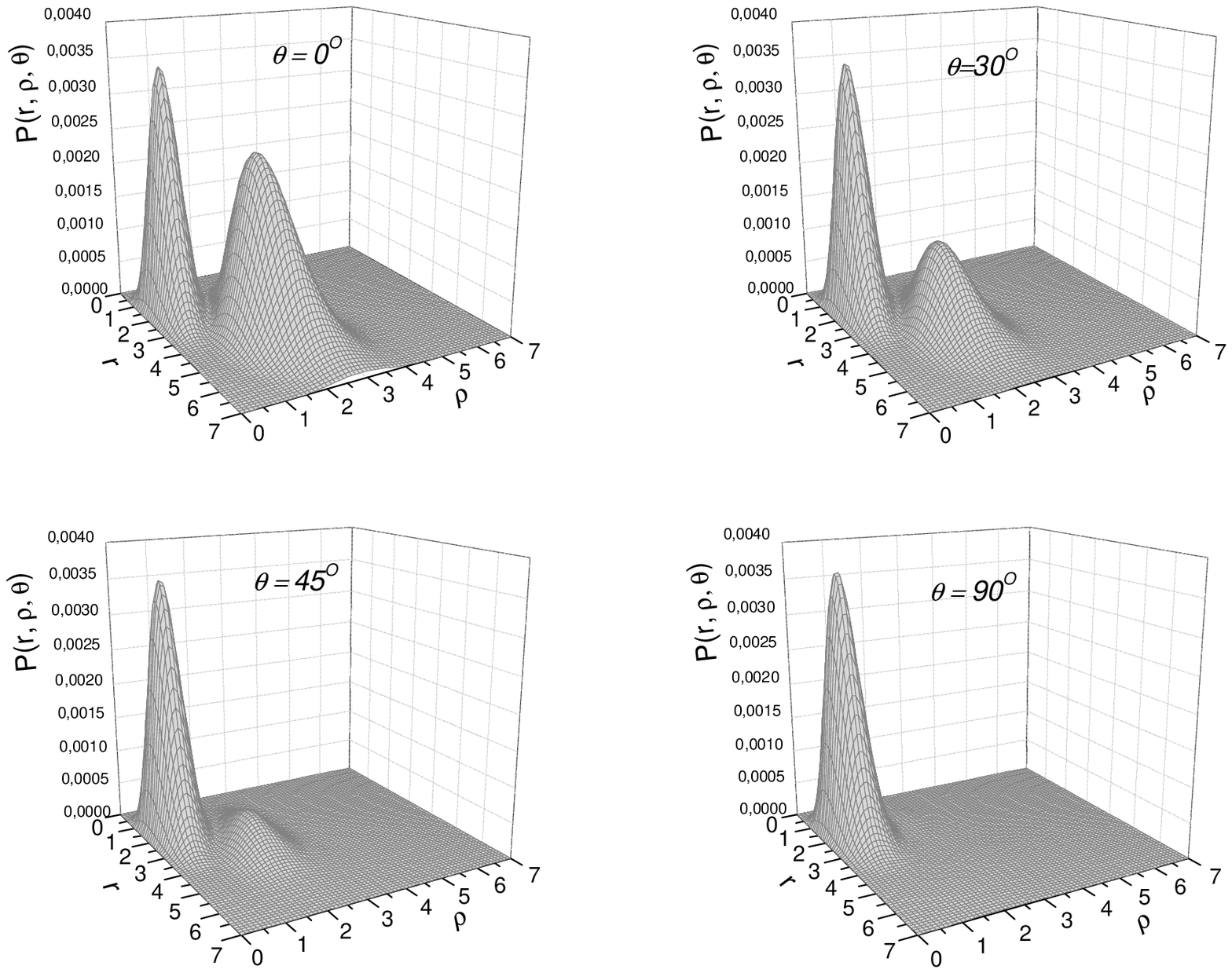}
\vskip-2mm\parbox{15.4cm}{\caption{Two configurations in the ground
state of $^{14}$C nucleus }}\vspace*{3mm}
\end{figure*}

\section{Momentum Distributions}

To complete the study of the structure functions of $^{14}$N
nucleus, we present the momentum distributions of $\alpha$-particles
and extra nucleons in this system within the five-particle
model.\,\,The momentum distribution $n_{i}\left(k\right)$ of the
$i$-th particle is known to be the density of the probability to
find this
particle with a definite momentum $k$,
\begin{equation}\label{E11}
n_{i}\left(k\right)=\left\langle\!\tilde{\Phi}\right|\delta
\left(\mathbf{k}-\left(\mathbf{k}_{i}-\mathbf{K}_{c.m.}\right)\right)
\left|\tilde{\Phi}\!\right\rangle\!,
\end{equation}
where $\tilde{\Phi}$ is the wave function of the system in the
momentum representation.\,\,The normalization of the momentum
distribution is $\int n_{i}\left(k\right)d\mathbf{k}=1$.\,\,The
momentum distribution $n_{i}\left(k\right)$ enables one, in
particular, to calculate the average kinetic energy of the $i$-th
particle:
\begin{equation}\label{E12}
\left\langle E_{i,{\rm
kin}}\right\rangle=\int\frac{k^{2}}{2m_{i}}n_{i}\left(k\right)d\mathbf{k}.
\end{equation}
Mainly due to the mass ratio between a nucleon and an
$\alpha$-particle, the extra nucleons move much more rapidly than
the $\alpha$-particles do.\,\,In particular, the average kinetic
energy of an extra proton in $^{14}$N nucleus is about $33.52$~MeV,
that of an extra neutron equals about $33.53$~MeV, while each of the
more slowly moving $\alpha$-particles has the kinetic energy of
about $5.79$~MeV.\,\,Similar values are typical of $^{14}$C and
$^{14}$O nuclei.\,\,In particular, the calculated kinetic energy of
each of the extra neutrons in $^{14}$C nucleus is about $32.66$~MeV,
while the same value for an $\alpha$-particle amounts about
$6.83$~MeV.\,\,For $^{14}$O nucleus, we have $31.77$~MeV for an
extra proton and
$6.62$~MeV for an $\alpha$-particle.\,\,%
\begin{figure}[h!]%
\vskip1mm
\includegraphics [width=\column] {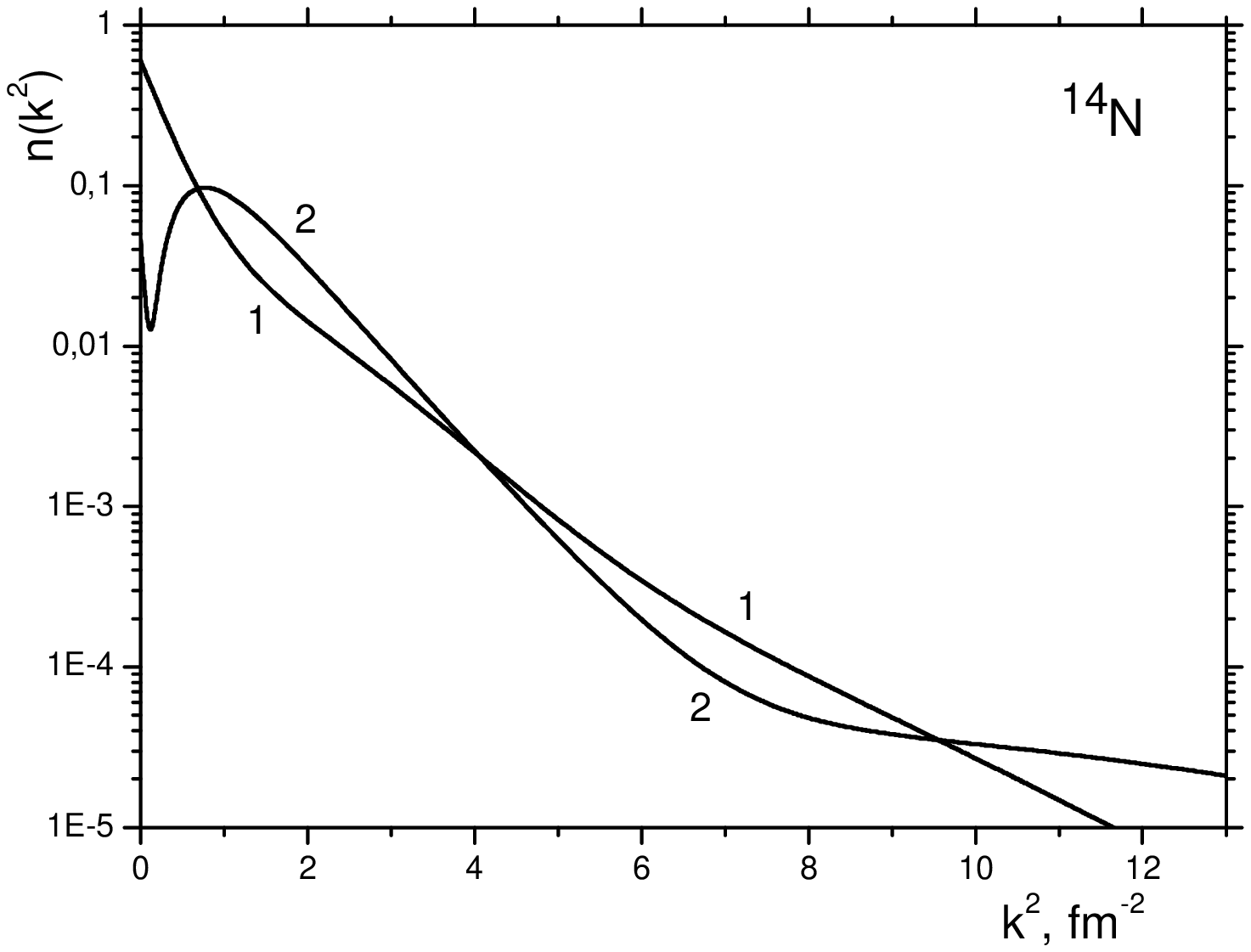}
\vskip-4mm\caption{Momentum distributions of an $\alpha$-particle
(curve \textit{1}) and an extra proton (curve \textit{2}) in
$^{14}$N nucleus}\vspace*{-5mm}
\end{figure}%
\begin{figure}[h!]%
\vskip1mm
\includegraphics [width=\column] {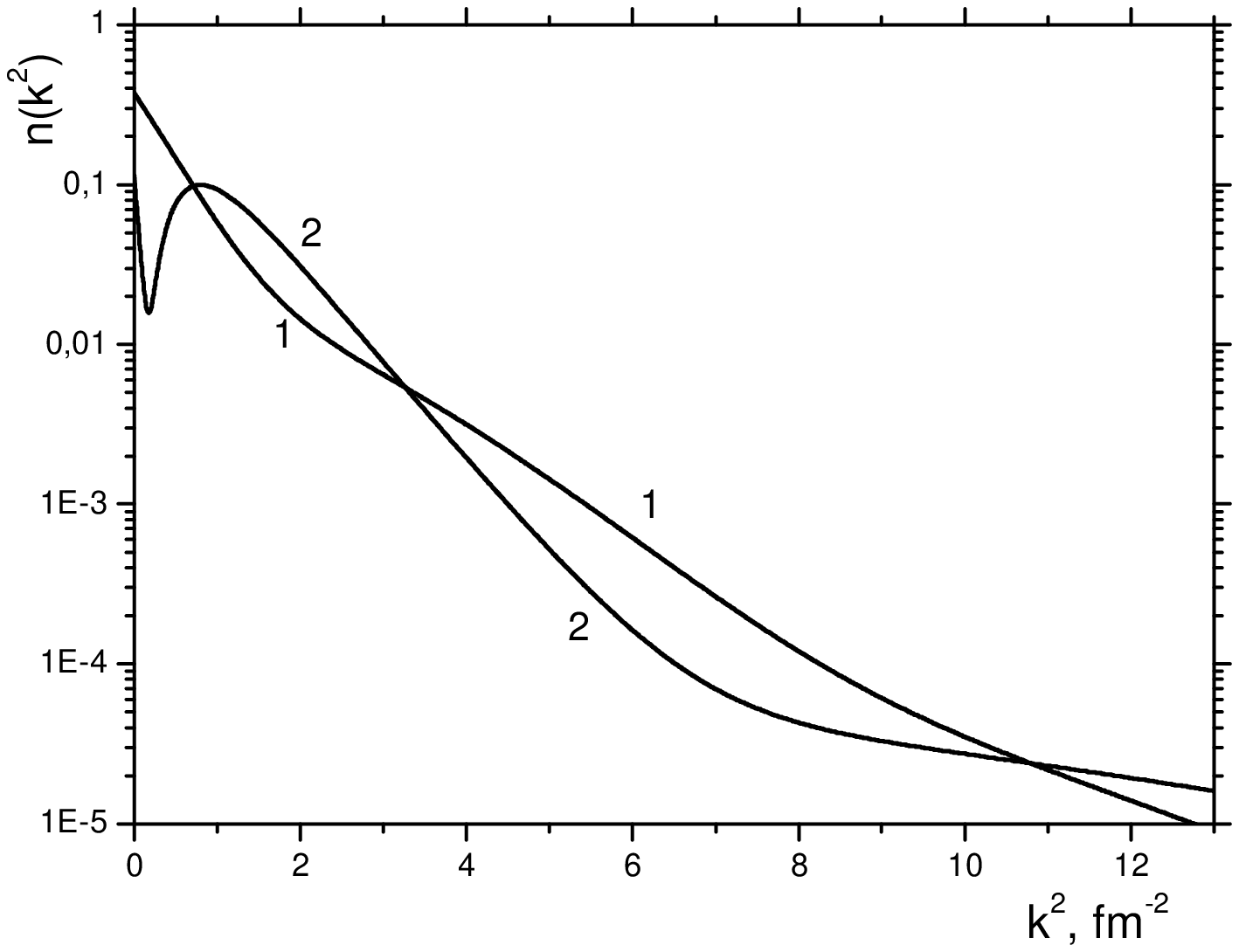}
\vskip-4mm\caption{Momentum distributions of an $\alpha$-particle
(curve \textit{1}) and an extra neutron (curve \textit{2}) in
$^{14}$C nucleus}\vspace*{-5mm}
\end{figure}%
The corresponding ratio of velocities is about $4.8$ for $^{14}$N
nucleus and about $4.4$ for $^{14}$C and $^{14}$O nuclei.\,\,This
means that the extra nucleons of the nuclei under consideration move
essentially faster than the heavier
$\alpha$-particles~do.

The momentum distributions of $\alpha$-particles, as well as those
of extra nucleons, are very similar for all the considered
nuclei.\,\,Especially, they are close for $^{14}$C and $^{14}$O
nuclei.\,\,That is why we present the profiles of the momentum
distributions only for $^{14}$N (Fig.~7) and $^{14}$C
(Fig.~8).\,\,In Fig.~7, curve \textit{1} corresponds to the momentum
distribution $n_{\alpha}\left(k\right)$ of an $\alpha$-particle, and
curve \textit{2} depicts $n_{p}\left(k\right)$ of an extra
proton.\,\,The momentum distribution of an extra neutron
$n_{n}\left(k\right)$ is not shown, because the corresponding curve
almost coincides with curve \textit{2}.\,\,Very similar (from the
qualitative point of view) profiles of the momentum distributions
are obtained for $^{14}$C and $^{14}$O nuclei (see Fig.~8 for
$^{14}$C; almost the same profiles could be depicted for $^{14}$O
nucleus).

The momentum distribution of $\alpha$-particles
$n_{\alpha}\left(k\right)$ is seen to be a monotonically decreasing
function, while $n_{p}\left(k\right)$ and $n_{n}\left(k\right)$ have
two maxima: at the zero momentum and at $k^{2}\simeq 1$
fm$^{-2}$.\,\,These two maxima correspond to two above-mentioned
configurations in the ground state of the nucleus.\,\,In a
configuration, where an extra nucleon is comparatively far from the
center of the nucleus, it moves comparatively slowly and makes a
contribution to the peak at very small $k^{2}$.\,\,If it is inside
$^{12}$C cluster (and this may occur in both spatial
configurations), its momentum is somewhat greater, and such momenta
make their contribution to the second maximum at $k^{2}\simeq 1$
fm$^{-2}$.\,\,At the same time, the heavier $\alpha$-particles
inside $^{12}$C cluster almost do not feel peculiarities of the
motion of extra nucleons.\,\,Thus, the influence of two different
spatial configurations of extra nucleons on the momentum
distribution of $\alpha$-particles is small due to both the mass
ratio and the comparatively large binding energy of $^{12}$C
cluster.\vspace*{-2mm}

\section{Conclusions}

To sum up, we note that the spatial structure of $^{14}$N nucleus
studied within the five-particle model is very similar to the
structure of the mirror nuclei $^{14}$C and $^{14}$O.\,\,Two
configurations in the ground-state wave functions of these nuclei
are revealed, where $^{12}$C cluster and the dinucleon subsystem
have the same centers of mass (first configuration, with a dinucleon
inside $^{12}$C cluster) or the shifted centers of mass (second
configuration, with one nucleon outside of $^{12}$C
cluster).\,\,These configurations manifest themselves, in
particular, in the density and momentum distributions.\,\,A similar
situation with two configurations in the ground state of the system
is inherent in some other light nuclei \cite{R1,R2,R3,R4,R5} with
two extra nucleons.

\vskip2mm

\textit{This work was supported in part by the Program of
Fundamental Research of the Division of Physics and Astronomy of the
NAS of Ukraine (project No.\,0117U000240).}




\rezume {Б.Є.\,Гринюк, Д.В.\,П'ятницький}{%
СТРУКТУРА ЯДРА $^{14}$N \\ У П'ЯТИКЛАСТЕРНІЙ МОДЕЛІ} {Досліджено
просторову структуру ядра $^{14}$N в рамках п'ятичастинкової моделі
(три $\alpha$-частинки і два нуклони). Розраховано енергію і
хвильову функцію основного стану цієї п'ятичастинкової системи на
основі варіаційного підходу з використанням гаусоїдних базисів.
Виявлено дві просторові конфігурації хвильової функції основного
стану. Проаналізовано розподіли густини, парні кореляційні функції і
імпульсні розподіли частинок в ядрі $^{14}$N та порівняно із
відповідними розподілами для дзеркальних ядер $^{14}$C~і~$^{14}$O.}

\end{document}